\documentclass[9pt,twocolumn,twoside]{opticajnl}
\journal{opticajournal} 

\setboolean{shortarticle}{true}

\usepackage{lineno}
\usepackage[utf8]{inputenc}

\title{Fully optoelectronic coherent THz spectroscopy}

\author[3]{François Parnet}
\author[2]{Francis Hindle}
\author[1]{Ulysse Mao}
\author[1]{Guillaume Ducournau}
\author[1]{Jean-François Lampin}
\author[1]{Sophie Eliet}
\author[3]{François Bondu}
\author[4]{Romain Peretti}
\author[2]{Gaël Mouret}
\author[2]{Daniele Fontanari}
\author[3]{Goulc'hen Loas}
\author[1,*]{Emilien Peytavit}

\affil[1]{IEMN, UMR CNRS 8520, Université de Lille, 59652, Villeneuve d’Ascq, France}
\affil[2]{Laboratoire de Physico-Chimie de l'Atmosphère, ULCO, Dunkerque 59140, France}
\affil[3]{Institut Foton, UMR 6082 / CNRS - Univ Rennes – INSA, France}
\affil[4]{INL, CNRS, École Centrale de Lyon, INSA Lyon, Université Claude Bernard Lyon 1, CPE Lyon, Villeurbanne, France.}

\affil[*]{emilien.peytavit@iemn.fr}

\begin{abstract}
Coherent terahertz spectroscopy of molecular transients has so far relied
on electronic sources whose bandwidth is limited to a fraction of their
center frequency. Photoconductive devices offer a complementary approach,
combining intrinsically broadband operation with coherent optoelectronic
generation and detection. We report free-induction-decay spectroscopy of
carbonyl sulfide at 0.29~THz in which, to our knowledge for the first
time, both the pulsed emitter and the coherent heterodyne receiver are
LT-GaAs photoconductors simultaneously pumped by a single free-running
dual-frequency Ti:Sa laser. The receiver operates within 15~dB of the
thermal noise limit, and the shared optical reference maintains phase
coherence over more than 1000 averaged acquisitions. The measured
transients agree with a time-domain Maxwell--Bloch model based on
HITRAN parameters. These results demonstrate the feasibility of fully
optoelectronic coherent THz spectroscopy and establish photoconductive
optoelectronic mixing as a practical route  toward frequency-agile, high-resolution coherent THz spectrometers over broad spectral ranges
\end{abstract}

\setboolean{displaycopyright}{false} 

\begin{document}
\maketitle

\section{Introduction}

Microwave chirped-pulse spectroscopy has become a reference technique for broadband, high-resolution rotational spectroscopy since its introduction by Brown \emph{et al.}~\cite{Brown2008,Abeysekera2014,Park2016}. Its success stems from the combination of coherent excitation and broadband detection of the molecular free induction decay (FID), enabling rapid acquisition of spectroscopic signatures with high sensitivity and absolute frequency accuracy. Continuous advances in microwave electronics have progressively extended this approach toward the millimeter (mm) and sub-millimeter (sub-mm) spectral regions~\cite{Park2011,Gerecht2011,Steber2012,Abeysekera2014,Zaleski2017,Hindle2017,Barnum2021,Hermanns2023}, where chirped-pulse spectrometers have become powerful tools for gas spectroscopy, molecular structure determination, and laboratory astrophysics.
\begin{figure}[!h]
\centering
\includegraphics[width=\linewidth]{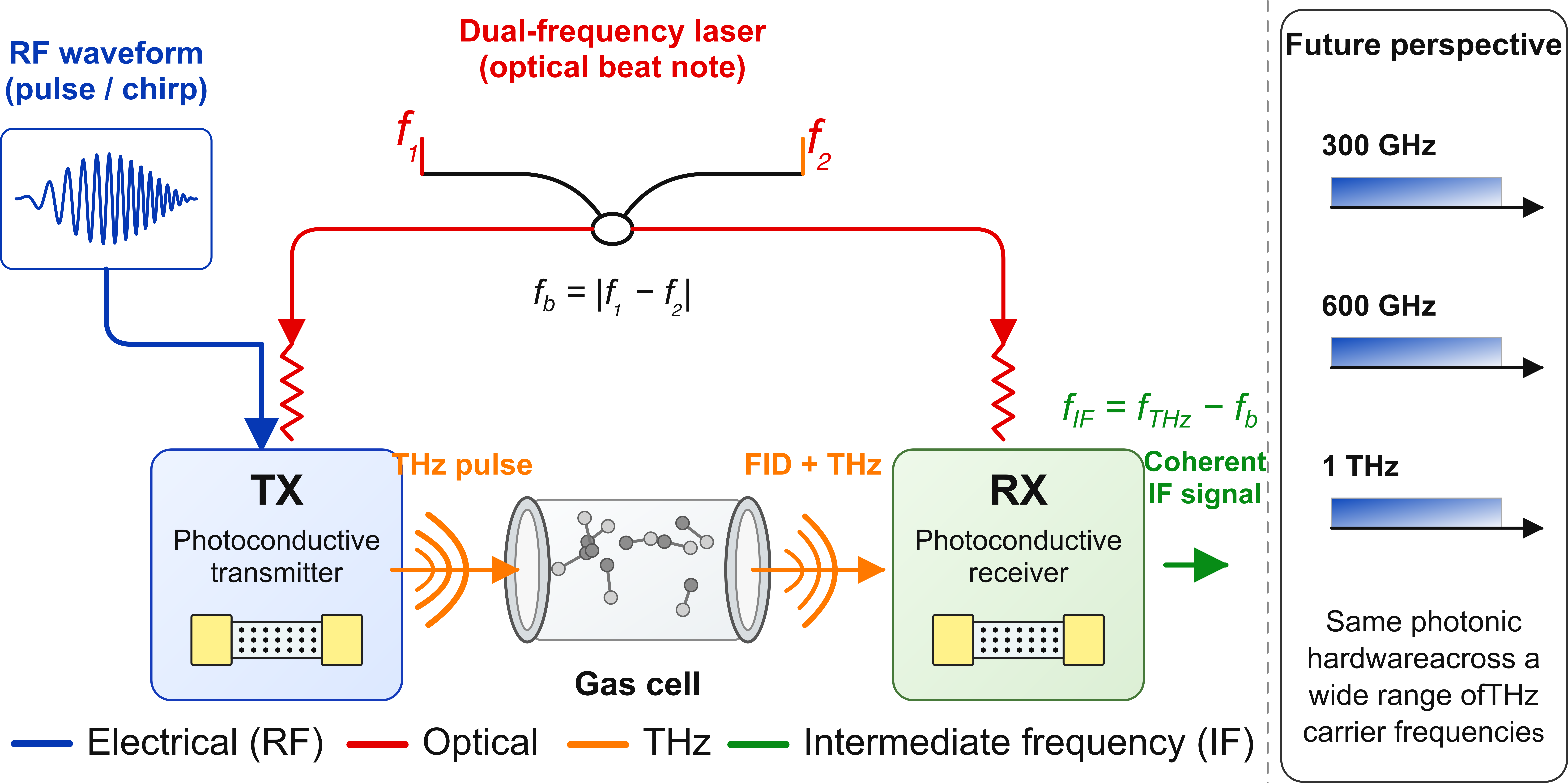}
\captionsetup{justification=justified,singlelinecheck=false}
\caption{Concept of photonic coherent THz spectroscopy. A dual-frequency laser provides an optical beat note at $f_b=|f_1-f_2|$ that simultaneously illuminates the photoconductive transmitter (TX) and receiver (RX). An arbitrary RF waveform, such as a pulse or chirp, drives the TX and generates a THz field at $f_{\mathrm{THz}}(t)=f_b+f_{\mathrm{RF}}(t)$. The molecular polarization builds up during excitation and subsequently emits a free-induction decay (FID), which is heterodyned in the RX by optoelectronic mixing to produce an intermediate-frequency signal at $f_{\mathrm{IF}}=f_{\mathrm{THz}}-f_b$. This architecture separates the optically defined THz carrier frequency from the electrically defined excitation waveform and could therefore enable the same photonic hardware to operate over a wide range of THz frequencies.}
\label{fig:concept}
\end{figure}
These instruments operate by coherently polarizing a molecular ensemble with a high-power chirped pulse and subsequently recording the resulting FID using phase-sensitive detection. Their performance strongly relies on modern high-speed electronics, including arbitrary waveform generators (AWGs), broadband frequency multipliers, and fast real-time oscilloscopes. These components enable the generation of repeatable, phase-coherent excitation pulses and efficient signal averaging over many acquisition cycles. However, extending this approach toward broader spectral coverage remains challenging. In the sub-mm region, electronic frequency multipliers, amplifiers, and mixers are typically limited to bandwidths of about $\pm20\%$ around their center frequency. Consequently, covering several hundreds of gigahertz, or ultimately the THz range, requires numerous frequency-specific components, resulting in increased system complexity and cost.

Photonic THz sources and detectors provide a natural route to overcome these limitations. In particular, optically driven photoconductors, commonly referred to as photomixers, can operate both as coherent emitters and receivers over exceptionally broad frequency ranges, typically from 100~GHz to well beyond 1~THz, as demonstrated, for instance, by photonic homodyne detection experiments above 1~THz~\cite{Verghese1998}. Their ultra-wide frequency tunability, combined with coherent generation and detection capabilities, makes them particularly attractive for extending chirped-pulse FID spectroscopy beyond the bandwidth limitations of purely electronic systems.

\begin{figure}[t]
\centering
\includegraphics[width=\linewidth]{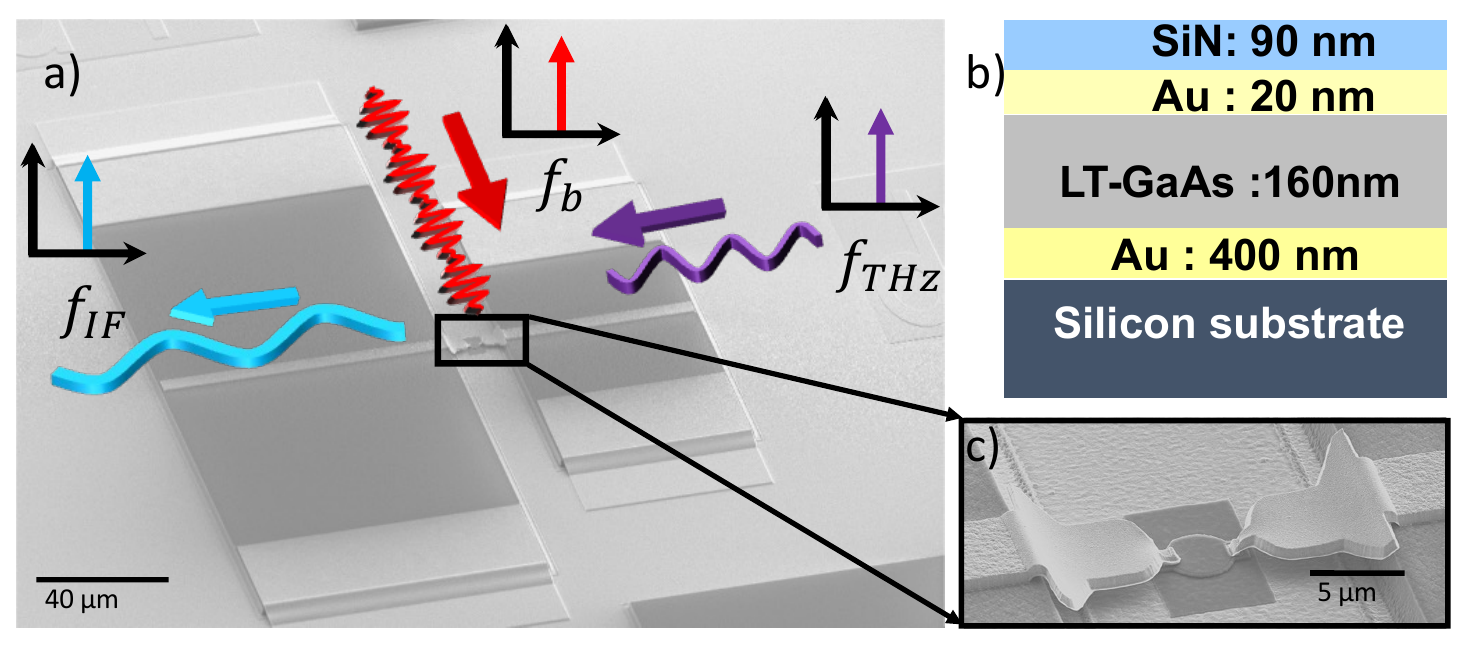}
\captionsetup{justification=justified,singlelinecheck=true}
\caption{Optoelectronic mixing receiver based on an optical-cavity-enhanced LT-GaAs photoconductor. 
(a) Scanning electron microscopy image of the photoconductor integrated with 50~$\Omega$ contact pads for input/output coupling via coplanar probes. The arrows indicate the optical beat note at frequency $f_b$ illuminating the photoconductor, the incident THz field at frequency $f_{\mathrm{THz}}$ coupled through the THz access, and the down-converted intermediate-frequency signal at $f_{\mathrm{IF}}=f_{\mathrm{THz}}-f_b$ extracted through the IF access.
(b) Layer structure of the photodetector; the layer thicknesses are designed to achieve peak optical absorption around 780~nm. 
(c) Close-up view of the device showing the intermediate-frequency (IF) and THz access.}
\label{fig:device}
\end{figure}
Unlike photodiodes, their responsivity can be
controlled by an applied electric field, enabling both pulsed THz
generation driven by an arbitrary RF waveform and coherent detection by
optoelectronic mixing. As illustrated in Fig.~\ref{fig:concept}, this
naturally separates the definition of the THz carrier frequency, fixed
by the optical beat note, from the excitation waveform, generated
electronically. In principle, the same photonic transmitter--receiver
pair could therefore implement pulse or chirped coherent spectroscopy
over a wide range of THz carrier frequencies without changing the active
devices.

Gas-phase THz time-domain spectroscopy (THz-TDS) has long demonstrated
the ability of photonic systems to perform broadband measurements of
molecular free-induction signals. In conventional THz-TDS, however, the
sub-picosecond excitation impulsively creates the molecular coherence,
without allowing the polarization to build up toward its steady-state
value, while the spectral resolution remains limited by the accessible
temporal observation window~\cite{BIGOURD20083111,Harde:91}. By
contrast, the present approach relies on narrowband coherent excitation,
allowing the molecular polarization to build up over several coherence
times before its free decay is recorded while retaining high spectral
resolution. Continuous-wave THz spectroscopy based on photomixing has
likewise been successfully applied to high-resolution gas spectroscopy,
both in homodyne and heterodyne configurations
\cite{Hindle2009,HINDLE2008262,Bigourd:06}.

The practical implementation of such a photonic coherent spectrometer,
however, requires combining several capabilities that have so far been
demonstrated separately: efficient THz generation, a low-noise coherent
receiver, and an optical beat note sufficiently stable to allow coherent
averaging of weak molecular transients. In this work, we demonstrate a
proof-of-principle experiment of fully optoelectronic coherent FID
spectroscopy by combining two devices previously developed by the
authors: (i) a low-temperature-grown (LT) GaAs Fabry--Perot cavity
photoconductor exhibiting conversion losses as low as 27~dB as a THz
receiver~\cite{peytavit2013c} and milliwatt-level THz emission up to
325~GHz~\cite{Peytavit2013}; and (ii) a dual-frequency Ti:Sa laser
providing an optical beat note with a frequency drift below 100~kHz over
more than 10~s and a linewidth narrower than 20~kHz over a 40~ms
measurement time~\cite{Loas2014}. Driven by this common optical
reference, the transmitter and receiver naturally remain phase coherent,
enabling coherent averaging of transient molecular responses.

After characterizing the combined optoelectronic performance of these
building blocks, we demonstrate coherent THz spectroscopy by recording
the free-induction decay of carbonyl sulfide (OCS) molecules excited by
a $\sim0.29$~THz pulse resonant with the
$J=24\leftarrow23$ rotational transition. The experimental transients
are quantitatively reproduced by a time-domain Maxwell--Bloch model
including Doppler broadening and collisional dephasing, thereby
establishing the feasibility of fully photonic coherent THz
spectroscopy.

\section{Experimental Results}

\begin{figure}[!t]
\centering
\includegraphics[width=\linewidth]{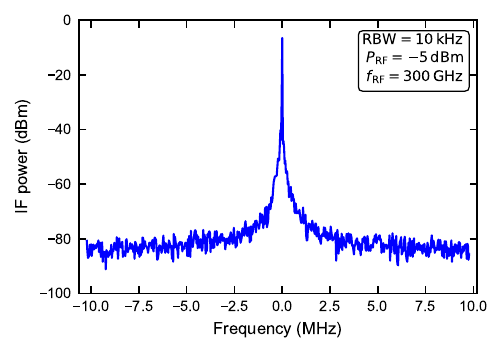}
\captionsetup{justification=justified,singlelinecheck=true}
\caption{Intermediate-frequency (IF) spectrum measured for an input RF
power of $-5$~dBm at 300~GHz, with the optical beat note tuned near
301~GHz (IF $\approx1$~GHz). The IF peak corresponds to a conversion
loss of approximately 30~dB [see Supplement~1 (Section~1)]; the
displayed noise floor illustrates the sensitivity level of the optoelectronic
heterodyne receiver}
\label{fig:ifspectrum}
\end{figure}

\begin{figure*}[t]
\centering
\includegraphics[width=\textwidth]{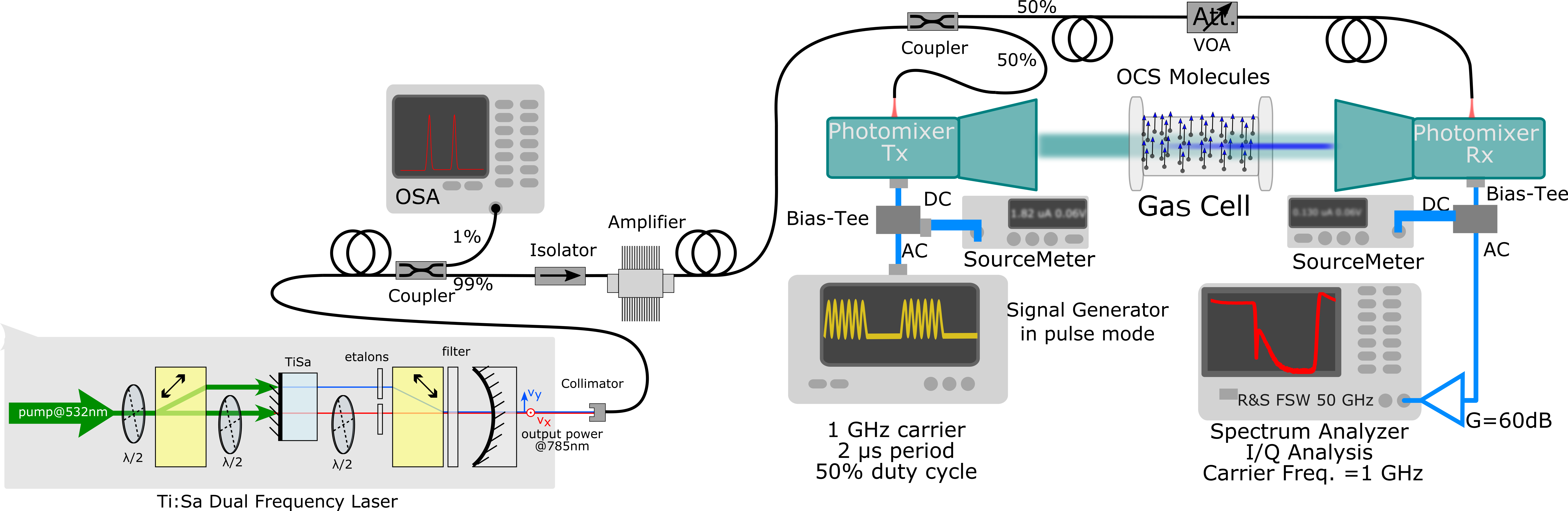}
\captionsetup{justification=justified,singlelinecheck=true}
\caption{Experimental setup of the optoelectronic FID experiment. THz pulses are generated by a biased photoconductor and launched through a coplanar probe and horn antenna into a gas cell. The dual-frequency 785~nm Ti:Sa laser features a 3~mm cavity incorporating a 1.7$^\circ$ active wedge, a BBO beam displacer providing a 1 mm beam separation for dual polarization, a 25~$\mu$m etalon (40\% reflectivity, 10~$\mu$rad wedge), and a 10~nm bandpass filter. This configuration ensures single-frequency operation of each of the two orthogonally polarized modes oscillating simultaneously in this bi-mode cavity, with a frequency drift of the resulting beat note below a few hundred kHz and a linewidth narrower than 20~kHz. The FID signal is detected using a heterodyne mixer identical to the transmitter (TX). The intermediate-frequency (IF) signal is measured in I/Q mode using a spectrum analyzer, with a 1~GHz carrier referenced to the TX signal generator.}
\label{fig:setup}
\end{figure*}

Figure~\ref{fig:device} illustrates the optoelectronic heterodyne receiver based on an optical-cavity-enhanced low-temperature-grown (LT) GaAs photoconductor. 
The heterodyne detection performance of the optoelectronic mixer was
evaluated using the configuration reported in Ref.~\cite{peytavit2013c},
in which a calibrated 300~GHz signal is applied to the on-wafer
photoconductor, the optical local oscillator being now provided by the
dual-frequency Ti:Sa laser. The beat note was tuned near 301~GHz,
setting the intermediate frequency close to 1~GHz, and the IF signal
was recorded with a spectrum analyzer.

Figure~\ref{fig:ifspectrum} shows the IF spectrum measured for an
input RF power of $-5$~dBm at 300~GHz. After correction for the IF
gain and losses, the observed peak corresponds to an IF power of about
$-35$~dBm at the mixer output, yielding a conversion loss of
approximately 30~dB; the calibration procedure is detailed in
Supplement~1, Section~1. This value places the present 780~nm GaAs
photomixer among the best photoconductive THz mixers reported at this
frequency: it improves by more than 20~dB over conventional planar
receivers, which historically exhibited 50--60~dB losses, and reaches
the level of the most recent 1550~nm Fe-doped InGaAs plasmonic and
waveguide-integrated devices~\cite{Deumer2025,Tannoury2023}.

In the same configuration, the displayed noise floor is $-90$~dBm in a
10~kHz resolution bandwidth, about 10~dB above the intrinsic analyzer
floor, corresponding to an input-referred noise density of
approximately $-160$~dBm/Hz, within 15~dB of the room-temperature
thermal noise limit (see Supplement~1, Section~1). The measurement is
therefore limited by the optoelectronic mixer and IF chain rather than
by the spectrum analyzer itself, with no need to invoke a dominant RIN
contribution from the optical beat note.

Photomixing emission from LT-GaAs photoconductors has already been
extensively characterized~\cite{Peytavit2013,Peytavit2011a}; the present setup
provides about $100~\mu$W of THz power at 300 GHz, comparable to
state-of-the-art photonic THz sources~\cite{Nagatsuma2016}. Together
with the state-of-the-art heterodyne sensitivity demonstrated above,
this raises the question of whether a fully photomixed coherent THz
system can detect the weak free induction decay of a dilute molecular
gas. We address this question using carbonyl sulfide (OCS) as a
benchmark molecule.

Figure~\ref{fig:setup} presents the experimental configuration of the optoelectronic free induction decay (FID) spectroscopy experiment. A train of coherent THz excitation pulses is generated by applying a pulsed RF bias to a low-temperature-grown GaAs photoconductor illuminated by the dual-frequency Ti:Sa laser. The optical beat note at frequency $f_{\mathrm{beat}}$ modulates the photocarrier density, while the applied RF bias at frequency $f_{\mathrm{RF}}$ mixes with this optical modulation. As a result, the emitted THz radiation contains frequency components at $f_{\mathrm{THz}} = f_{\mathrm{beat}} + f_{\mathrm{RF}}$ as well as higher-order harmonics. In the present experiment, the carrier is tuned close to the OCS rotational transition at 291.839~GHz.
The generated THz pulses are coupled into a narrowband (220--325~GHz) coplanar waveguide probe, radiated into free space using a horn antenna, and transmitted through a gas cell containing OCS. The pulsed excitation induces a transient macroscopic molecular polarization, which re-emits coherently after the excitation pulse, giving rise to the free induction decay signal.
Detection is performed using a second photomixer operated as a coherent heterodyne receiver. Both transmitter and receiver are illuminated by the same dual-frequency laser and therefore share a common optical beat note and phase reference. In the receiver, the incident THz field acts as an effective AC bias across the photoconductor and mixes with the optical beat modulation. This down-conversion process generates an intermediate frequency given by $f_{\mathrm{IF}} = f_{\mathrm{THz}} - f_{\mathrm{beat}}$ so that, for the fundamental mixing order, one naturally obtains $f_{\mathrm{IF}} = f_{\mathrm{RF}}$. The resulting IF signal is amplified by 65~dB and recorded with a spectrum analyzer (R\&S FSW) operated in I/Q mode. Although the heterodyne scheme preserves the field quadratures, the measurements
reported below use only the amplitude reconstructed from them, $|S_{\mathrm{IF}}|=(I^2+Q^2)^{1/2}$.

\begin{figure}[!t]
\centering
\includegraphics[width=\linewidth]{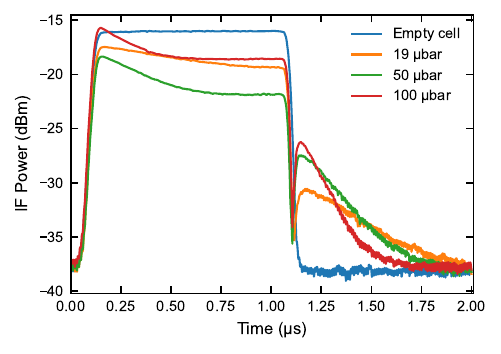}
\captionsetup{justification=justified,singlelinecheck=true}
\caption{Experimental time-resolved IF power over a single modulation period (0--2~$\mu$s), recentered at the onset of the excitation pulse. Traces correspond to the empty cell (blue) and OCS pressures of 19, 50, and 100~$\mu$bar. The plotted quantity is the power envelope of the detected field obtained from the I/Q data; the underlying 1~GHz IF carrier oscillations are not resolved on this time scale. During the excitation pulse the trace reflects the absorption of the incident field by the gas, while the signal observed after the pulse corresponds to the coherent emission (FID) of the sample.}
\label{fig:fid}
\end{figure}

Typical FID signals are shown in Fig.~\ref{fig:fid} for different OCS pressures. Following the excitation pulse, a coherent reemission is observed,
whose pressure dependence reflects the trade-off between molecular
density and collisional dephasing: from 19 to 100~$\mu$bar the FID
amplitude increases while its decay accelerates. Since the acquisition
trigger and the IF carrier are both referenced to the TX signal
generator, 1000 successive I/Q traces can be averaged coherently by
the spectrum analyzer, significantly improving the signal-to-noise
ratio.

To quantitatively interpret these transient responses and separate absorption, dispersion, Doppler dephasing, and collisional effects, a time-domain coherent model based on the optical Bloch formalism and Doppler averaging is introduced in the following section.

\section{Time-domain coherent model}

\begin{figure}[t!]
\centering
\includegraphics[width=\linewidth]{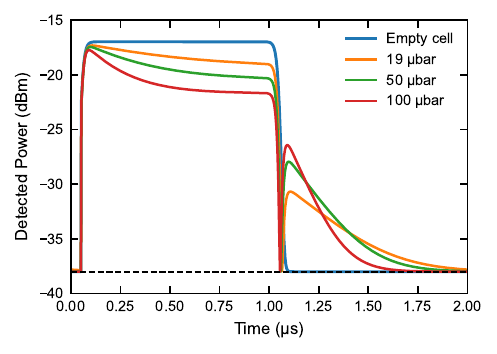}
\captionsetup{justification=justified,singlelinecheck=true}
\caption{Simulated time-domain detected power over a single modulation period (0--2~$\mu$s), obtained from the coherent Maxwell--Bloch model described in this section, i.e. the numerical integration of the polarization equation~(\ref{eq:polar}) summed over the Doppler velocity classes [Eq.~(\ref{eq:doppler})] and inserted in the thin-sample propagation relation~(\ref{eq:thinsample}). Traces correspond to OCS pressures of 19, 50, and 100~$\mu$bar; the empty cell response is shown for reference. An instrumental noise floor identical to the experiment is added for direct comparison.}
\label{fig:fid_model}
\end{figure}

The molecular response is modeled in the linear regime of the optical
Bloch equations for a weakly driven rotational transition
~\cite{AllenEberly1987}. In the following, all field and polarization
amplitudes denote slowly varying complex envelopes in the rotating
frame. A rigorous derivation of the reduced model used below, including
the envelope conventions and the validity conditions of the
approximations, is provided in Supplement~1, Section~2.

In the linear regime, the population difference remains close to its
thermal equilibrium value and the dynamics reduces to that of the
molecular coherence. Equivalently, the macroscopic polarization obeys
\begin{equation}
\frac{dP}{dt}
=
\left(i\Delta\omega-\frac{1}{T_2}\right)P
+
\kappa E(t),
\label{eq:polar}
\end{equation}
where $\Delta\omega=\omega_0-\omega$ is the detuning between the
molecular transition and the excitation field, $T_2$ is the coherence
time, and $\kappa$ is the molecular coupling coefficient.

The detected field is obtained by combining this molecular response
with the slowly varying Maxwell equation. In the optically thin-sample
limit, the transmitted envelope can be written as
\begin{equation}
E_{\mathrm{out}}(t)
=
E_{\mathrm{in}}(t)
+
\beta P(t),
\label{eq:thinsample}
\end{equation}
where $\beta$ is an effective complex propagation and detection
coefficient. Equation~(\ref{eq:thinsample}) emphasizes that the
detected signal is the coherent superposition of the incident field and
the field radiated by the molecular polarization.

For an ideal square excitation pulse, Eq.~(\ref{eq:polar}) predicts the
build-up of the driven polarization toward a steady-state value,
followed after switch-off by a free-induction decay on the time scale
$T_2$, with an additional Gaussian dephasing term when Doppler
broadening is included. In the experiment, however, the excitation pulse
has finite rise and fall times of a few nanoseconds. During these
switching edges, the incident field can interfere strongly with the
molecular emission; in particular, the residual molecular polarization
can transiently radiate with an amplitude and phase such that the total
detected field nearly vanishes before the free-induction decay
dominates.

To account for this effect and for Doppler broadening, the polarization
dynamics is solved numerically in the time domain. Equation
~(\ref{eq:polar}) is integrated independently for each Doppler class
$j$, with detuning $\Delta\omega-2\pi\nu_j$ and statistical weight
$g_j$, and the total polarization is calculated as
\begin{equation}
P(t)=\sum_j g_j P_j(t).
\label{eq:doppler}
\end{equation}

The simulation parameters are derived from the HITRAN database for the
OCS $J=24\leftarrow23$ transition at 296~K. The collision-limited
coherence time and Doppler broadening are entirely determined from the
spectroscopic parameters, while an effective optical depth is adjusted
for each pressure to account for small variations in excitation
detuning and experimental coupling (Supplement~1, Sections~3 and~4).

Overall, the simulations reproduce the main features of the experimental
FID traces, including the pressure-dependent decay and the transient
interference pattern following the excitation pulse. On the averaged traces, the FID peak lies 8--12~dB above the instrumental noise floor. Since the detected power scales as $p^2$, comparison of the peak signal-to-noise ratio with the measured baseline fluctuations yields a $3\sigma$ minimum detectable pressure of approximately $4~\mu\mathrm{bar}$. Applying a matched filter to the recorded power envelope further improves the detection limit to approximately $1.3~\mu\mathrm{bar}$ (see Supplement~1, Section~5).
The main discrepancies concern the absolute signal levels.
Experimentally, contrary to what would be expected, the detected level immediately after the rising edge,
before molecular polarization is established, differs slightly between
measurements, and noticeable differences are also observed in the
amplitude of the bump associated with the FID. The experimental variations are most likely due to small fluctuations of the emitted THz power and to changes in the conversion efficiency of
the optoelectronic mixer, which is sensitive to the optical coupling
between the lensed fiber illumination and the photoconductor active
area. In addition, a slow drift of the dual-frequency laser may slightly
shift the carrier frequency from line center between acquisitions,
which also affects the observed amplitudes after averaging 1000 I/Q
traces.

In conclusion, we have demonstrated coherent THz spectroscopy through free induction decay detection using an optoelectronic heterodyne receiver driven by a dual-frequency Ti:Sa laser. The excellent agreement between the experimental results and the time-domain model confirms the coherent nature of the detected molecular response and validates the proposed approach. 

Beyond this proof of principle, the present work establishes optoelectronic mixing as a promising platform for compact, frequency-agile THz spectrometers. Because the excitation waveform is entirely defined by the RF bias applied to the photoconductor, arbitrary pulse shapes, including chirped excitation pulses, can in principle, be synthesized without modifying the optical source. This opens the way to extending chirped-pulse spectroscopy to spectral regions that remain difficult or inaccessible using conventional electronic multiplier-based architectures.

Such an approach would enable high-resolution Doppler-limited spectroscopy over previously unexplored THz bands, providing access to molecular transitions that are currently difficult to investigate and thereby improving spectroscopic databases and molecular models. More broadly, the combination of ultra-broad spectral coverage, coherent detection, and microsecond time resolution offers attractive prospects for time-resolved studies of complex gas-phase reaction kinetics. Finally, owing to the relatively low scattering of THz radiation by aerosols and suspended particles, optoelectronic THz chirped-pulse spectroscopy could become a valuable tool for investigating heterogeneous chemical processes, where the interactions between gas-phase species and particles, as well as the catalytic role of particulate matter, remain only partially understood\cite{Bigourd:06}.

\section*{Funding}
Agence Nationale de la Recherche (ANR) (PHENIX, ANR-15-CE24-0004; OSCAR, ANR-15-CE29-0017); 
RENATECH network (French national nanofabrication network).
RF-Net network (ANR-22-PEFT-0011).
\section*{Disclosures}
The authors declare no conflicts of interest.
\section*{Data availability}
Data underlying the results presented in this paper are not publicly
available at this time but may be obtained from the authors upon
reasonable request.
\section*{Supplemental document}
See Supplement 1 for supporting content.
\bibliography{sample}

\end{document}


\maketitle

\section{Heterodyne mixer measurement}
\label{sec:supp_mixer}
To characterize the conversion loss of the receiver, a calibrated RF
signal at 300~GHz was generated using a frequency-multiplier chain and
applied to the on-wafer photoconductor through a waveguide-coupled
coplanar probe, following the configuration of E.~Peytavit \textit{et
al.}, Appl. Phys. Lett. \textbf{103}, 201107 (2013), cited in the main
text. 
The frequency difference between the two modes of the
dual-frequency Ti:Sa laser was adjusted to approximately 301~GHz,
providing the optical local oscillator and setting the intermediate
frequency (IF) close to 1~GHz. The IF signal was extracted through a
second coplanar probe connected to a coaxial line, amplified by a
30~dB low-noise amplifier, and recorded with a spectrum analyzer
(Rohde \& Schwarz FSW).

The RF input power quoted in the main text, $P_{\mathrm{RF}}=-5$~dBm,
is corrected for the insertion losses of the probes, cables, and other
components of the RF measurement chain. For this input power, the IF
peak reads $-6.5$~dBm on the analyzer trace. Subtracting the 30~dB
gain of the IF amplifier and adding back the $\approx2$~dB of
insertion loss between the mixer output and the amplifier input yields
\begin{equation}
P_{\mathrm{IF}} \approx -6.5 - 30 + 2 = -34.5 \approx -35~\mathrm{dBm}
\end{equation}
at the mixer output. The conversion loss then follows as
\begin{equation}
L_c = P_{\mathrm{RF}} - P_{\mathrm{IF}} \approx -5-(-35) = 30~\mathrm{dB}.
\end{equation}

The effective noise density was estimated from the noise floor measured
on the same analyzer trace. In the configuration used for Fig.~2 of the
main text, the displayed noise floor is about $-90$~dBm in a 10~kHz
resolution bandwidth. This corresponds to a noise power density at the
analyzer input of
\begin{equation}
N_{\mathrm{SA}}
= -90 - 10\log_{10}(10^4)
= -130~\mathrm{dBm/Hz}.
\end{equation}
Referring this value back to the input of the IF amplifier by
subtracting the 30~dB amplifier gain gives
\begin{equation}
N_{\mathrm{in}}
\approx -130 - 30
= -160~\mathrm{dBm/Hz}.
\end{equation}
This effective input-referred noise density lies about 14~dB above
the room-temperature thermal noise density,
$k_{\mathrm{B}}T \simeq -174~\mathrm{dBm/Hz}$. The intrinsic spectrum
analyzer floor measured under identical settings, but without the
optoelectronic receiver chain, was approximately 10~dB lower than the
displayed floor in Fig.~2 of the main text, confirming that the
measurement is not limited by the analyzer itself.

\section{Molecular response and propagation model}
\label{sec:supp_model}

We summarize here the derivation of the molecular response and
propagation model used in the main text. The molecule is described as
an effective two-level rotational transition of angular frequency
$\omega_0$, driven by a quasi-monochromatic THz field of carrier
angular frequency $\omega$. We use slowly varying positive-frequency
amplitudes, defined by
\begin{equation}
E_{\mathrm{real}}(t)
=
E(t)e^{-i\omega t}
+
\mathrm{c.c.},
\label{eq:supp_E_positive}
\end{equation}
where $E(t)$ is the complex envelope used in the main text. The
molecular coherence is written in the same rotating frame,
\begin{equation}
\rho_{12}^{\mathrm{lab}}(t)
=
\tilde{\rho}_{12}(t)e^{-i\omega t},
\end{equation}
where $\tilde{\rho}_{12}$ is the slowly varying coherence. With the
population difference defined as
\begin{equation}
w=\rho_{11}-\rho_{22},
\end{equation}
the optical Bloch equations under the rotating-wave approximation give
\begin{equation}
\frac{d\tilde{\rho}_{12}}{dt}
=
\left(i\Delta\omega-\frac{1}{T_2}\right)\tilde{\rho}_{12}
+
i\frac{\mu}{\hbar}E(t)w ,
\label{eq:supp_bloch_general}
\end{equation}
where $\Delta\omega=\omega_0-\omega$ is the angular-frequency detuning,
$\mu$ is the transition dipole moment, and $T_2$ is the coherence
dephasing time. 

In the weak-saturation regime, the population difference remains close
to its thermal equilibrium value, $w\simeq w_{\mathrm{eq}}$. This
requires the saturation parameter to remain small,
\begin{equation}
\Omega^2 T_1T_2 \ll 1,
\qquad
\Omega(t)=\frac{2\mu E(t)}{\hbar},
\label{eq:supp_saturation}
\end{equation}
where $T_1$ is the population relaxation time and $\Omega$ is the Rabi
frequency in the present envelope convention. Equation
~\eqref{eq:supp_bloch_general} then reduces to the linear coherence
equation
\begin{equation}
\frac{d\tilde{\rho}_{12}}{dt}
=
\left(i\Delta\omega-\frac{1}{T_2}\right)\tilde{\rho}_{12}
+
i\frac{\mu}{\hbar}E(t)w_{\mathrm{eq}} .
\label{eq:supp_bloch_linear}
\end{equation}

The real macroscopic polarization is written with the same envelope
convention,
\begin{equation}
P_{\mathrm{real}}(t)
=
P(t)e^{-i\omega t}
+
\mathrm{c.c.}
\label{eq:supp_P_positive}
\end{equation}
For a molecular number density $N$, the slowly varying polarization
amplitude is
\begin{equation}
P(t)=N\mu\tilde{\rho}_{12}(t).
\label{eq:supp_macroP}
\end{equation}
Combining Eqs.~\eqref{eq:supp_bloch_linear} and
\eqref{eq:supp_macroP} gives
\begin{equation}
\frac{dP}{dt}
=
\left(i\Delta\omega-\frac{1}{T_2}\right)P
+
\kappa E(t),
\label{eq:supp_polar}
\end{equation}
with
\begin{equation}
\kappa=i\frac{N\mu^2}{\hbar}w_{\mathrm{eq}} .
\label{eq:supp_kappa}
\end{equation}

When Doppler broadening is included, the detuning is replaced by the
detuning experienced by each velocity class. We denote by $P_j(t)$ the
polarization response calculated for a Doppler class $j$ with frequency
shift $\nu_j$, and by $g_j$ its statistical weight, with
$\sum_j g_j=1$. Each class obeys
\begin{equation}
\frac{dP_j}{dt}
=
\left[
i\left(\Delta\omega-2\pi\nu_j\right)
-\frac{1}{T_2}
\right]
P_j
+
\kappa E(t).
\label{eq:supp_doppler_class}
\end{equation}
The total polarization is then obtained as
\begin{equation}
P(t)=\sum_j g_j P_j(t).
\label{eq:supp_doppler_sum}
\end{equation}
In the simulations reported in the main text, the weights $g_j$
reproduce the one-dimensional Maxwell--Boltzmann velocity distribution
projected along the THz propagation axis (see
Section~\ref{sec:supp_parameters}).
We now relate the molecular polarization to the detected field. We
consider a plane wave propagating along $z$ with the same carrier
frequency $\omega$,
\begin{align}
E_{\mathrm{real}}(z,t)
&=
E(z,t)e^{i(kz-\omega t)}
+
\mathrm{c.c.},
\\
P_{\mathrm{real}}(z,t)
&=
P(z,t)e^{i(kz-\omega t)}
+
\mathrm{c.c.},
\end{align}
where $k=\omega/c$. The envelopes $E$ and $P$ are assumed to vary
slowly compared with the carrier oscillation. In one dimension,
Maxwell's equations give the driven wave equation
\begin{equation}
\frac{\partial^2 E_{\mathrm{real}}}{\partial z^2}
-
\frac{1}{c^2}
\frac{\partial^2 E_{\mathrm{real}}}{\partial t^2}
=
\mu_0
\frac{\partial^2 P_{\mathrm{real}}}{\partial t^2}.
\label{eq:supp_wave}
\end{equation}
Substituting the complex envelopes into Eq.~\eqref{eq:supp_wave}, using
$k=\omega/c$, and applying the slowly varying envelope approximation
gives
\begin{equation}
2ik\frac{\partial E}{\partial z}
+
\frac{2i\omega}{c^2}
\frac{\partial E}{\partial t}
=
-\mu_0\omega^2 P.
\end{equation}
Using $\mu_0c^2=1/\varepsilon_0$, one obtains
\begin{equation}
\frac{\partial E}{\partial z}
+
\frac{1}{c}
\frac{\partial E}{\partial t}
=
i\frac{\omega}{2\varepsilon_0 c}P.
\label{eq:supp_svea_lab}
\end{equation}
In the retarded-time frame $\tau=t-z/c$, this becomes
\begin{equation}
\frac{\partial E}{\partial z}
=
i\frac{\omega}{2\varepsilon_0 c}P,
\label{eq:supp_svea_retarded}
\end{equation}
where the derivative is taken at fixed retarded time.

For an optically thin sample of length $L$, Eq.
~\eqref{eq:supp_svea_retarded} can be integrated by neglecting the
variation of the polarization over the propagation distance. Denoting
by $E_{\mathrm{in}}(t)=E(0,t)$ and
$E_{\mathrm{out}}(t)=E(L,t)$ the incident and transmitted envelopes,
one obtains
\begin{equation}
E_{\mathrm{out}}(t)
=
E_{\mathrm{in}}(t)
+
\beta P(t),
\label{eq:supp_thin}
\end{equation}
with
\begin{equation}
\beta=i\frac{\omega L}{2\varepsilon_0 c}.
\label{eq:supp_beta}
\end{equation}
This expression assumes that the molecular field remains a small
perturbation of the incident field,
$|\beta P|\ll |E_{\mathrm{in}}|$, and that propagation retardation
across the sample can be neglected on the relevant temporal scales,
namely $L/c\ll T_2$ and $L/c$ much shorter than the pulse rise and fall
times. In the experimental analysis, $\beta$ is treated as an effective
complex calibration parameter, so that it also accounts for propagation
losses, antenna coupling efficiency, and the complex gain of the
heterodyne detection chain.

\section{Spectroscopic parameters and line broadening}
\label{sec:supp_parameters}

\subsection{HITRAN parameters}

The spectroscopic parameters used in the simulations are derived from
the HITRAN database for the OCS rotational transition
$J=24\rightarrow23$ at $T=296$~K. The relevant parameters for the
isotopologue $^{16}$O$^{12}$C$^{32}$S are summarized in
Table~\ref{tab:supp_hitran}.

\begin{table}[h]
\centering
\begin{tabular}{lc}
\hline
Quantity & Value \\
\hline
Line position $\tilde{\nu}_0$ & $9.73472~\mathrm{cm^{-1}}$ \\
Line intensity $S$ & $1.027\times10^{-21}~\mathrm{cm\,molecule^{-1}}$ \\
Einstein coefficient $A$ & $7.24\times10^{-5}~\mathrm{s^{-1}}$ \\
Self broadening $\gamma_{\mathrm{self}}$ & $0.171~\mathrm{cm^{-1}\,atm^{-1}}$ \\
Lower-state energy $E''$ & $111.964~\mathrm{cm^{-1}}$ \\
\hline
\end{tabular}
\caption{HITRAN parameters used in the simulations.}
\label{tab:supp_hitran}
\end{table}

The transition frequency is obtained from
\begin{equation}
\nu_0=c\tilde{\nu}_0 \approx 291.839~\mathrm{GHz}.
\end{equation}

\subsection{Collisional broadening and coherence time}

The pressure-dependent Lorentz half-width (collisional, homogeneous
broadening) is obtained from the HITRAN self-broadening coefficient
\begin{equation}
\gamma_{\mathrm{cm^{-1}}}(p)=\gamma_{\mathrm{self}}\,p .
\end{equation}
Conversion to frequency units uses
\begin{equation}
\gamma_{\mathrm{Hz}} = c\,\gamma_{\mathrm{cm^{-1}}}.
\end{equation}
The collision-limited coherence time $T_2$ is then
\begin{equation}
T_2=\frac{1}{2\pi\gamma_{\mathrm{Hz}}}.
\end{equation}
The values derived from HITRAN are listed in
Table~\ref{tab:supp_T2hitran}.

\begin{table}[h]
\centering
\begin{tabular}{cc}
\hline
Pressure ($\mu$bar) & $T_2$ ($\mu$s) \\
\hline
19 & 1.6 \\
50 & 0.62 \\
100 & 0.31 \\
\hline
\end{tabular}
\caption{Collision-limited coherence times derived from HITRAN.}
\label{tab:supp_T2hitran}
\end{table}

\subsection{Doppler broadening}

Thermal motion produces a Gaussian distribution of molecular velocities
which translates into a Gaussian (inhomogeneous) Doppler frequency
distribution with standard deviation
\begin{equation}
\sigma_\nu=\nu_0\sqrt{\frac{k_B T}{mc^2}} .
\end{equation}
For OCS ($m\approx60$~amu) at $296$~K this gives
\begin{equation}
\sigma_\nu \approx 2\times10^5~\mathrm{Hz}\;=0.2~\mathrm{MHz},
\end{equation}
corresponding to a Doppler FWHM of approximately
\begin{equation}
\Delta\nu_D \approx 0.46~\mathrm{MHz}.
\end{equation}
The Doppler distribution is implemented numerically by discretizing the
velocity distribution into $N$ frequency classes of Doppler shifts
$\nu_j$ with statistical weights
$g_j\propto\exp\!\left(-\nu_j^2/2\sigma_\nu^2\right)$, normalized such
that $\sum_j g_j=1$, consistently with the notation of
Section~\ref{sec:supp_model}.

\subsection{On-resonance optical depth}

The on-resonance optical depth is estimated from the HITRAN line
intensity as
\begin{equation}
\mathrm{OD}_0 = N S L\, g(\nu_0),
\end{equation}
where $N$ is the molecular number density, $L$ the interaction length,
and $g(\nu)$ the normalized line shape, taken as a Voigt profile
combining the Gaussian Doppler contribution
($\Delta\nu_D\approx0.46$~MHz FWHM) and the Lorentzian collisional
contribution ($2\gamma_{\mathrm{Hz}}\approx0.2$~MHz FWHM at
20~$\mu$bar); at the lowest pressures the line shape is thus
essentially Doppler-limited.

For the present experimental conditions ($L\approx1$~m), the peak value
of the normalized Voigt profile is computed numerically from the real
part of the Faddeeva function, yielding
\begin{equation}
\mathrm{OD}_0^{\mathrm{HITRAN}}\approx 2.1,\; 3.5,\;\mathrm{and}\; 4.3
\end{equation}
at 19, 50, and 100~$\mu$bar, respectively. We note that the purely
Lorentzian estimate $g(\nu_0)=1/(\pi\gamma)$, which leads to the
pressure-independent value $\mathrm{OD}_0=SL/(\pi\gamma_{\mathrm{self}}
k_B T)\approx4.7$, is only valid in the collision-dominated regime; at
the pressures considered here, and in particular below the crossover
pressure of $\approx46~\mu$bar where the collisional width equals the
Doppler width, it overestimates the on-resonance optical depth.

\section{Numerical implementation}
\label{sec:supp_numerics}

The polarization dynamics is solved numerically in the time domain by
integrating the Doppler-class equation~\eqref{eq:supp_doppler_class}
with an explicit Euler scheme, the driving field being identified with
the incident envelope $E_{\mathrm{in}}(t)$ in the thin-sample limit of
Eq.~\eqref{eq:supp_thin}:
\begin{equation}
P_j^{(n+1)} =
P_j^{(n)} + \Delta t
\left\{
\left[
i\left(\Delta\omega-2\pi\nu_j\right)-\frac{1}{T_2}
\right]P_j^{(n)}
+
\kappa E_{\mathrm{in}}^{(n)}
\right\},
\label{eq:supp_euler}
\end{equation}
where $\nu_j$ and $g_j$ are the Doppler shift and the Gaussian
statistical weight of velocity class $j$ defined in
Sections~\ref{sec:supp_model} and~\ref{sec:supp_parameters}, and
$E_{\mathrm{in}}^{(n)}=E_{\mathrm{in}}(t_n)$ is the incident field
envelope sampled at time $t_n=n\,\Delta t$. The total polarization is
obtained from the weighted sum of Eq.~\eqref{eq:supp_doppler_sum}, and
the detected field from the input--output relation of
Eq.~\eqref{eq:supp_thin}. The coherence time $T_2$ entering the simulations is the
collision-limited value derived from the HITRAN self-broadening
coefficient (Table~\ref{tab:supp_T2hitran}).

The sampling interval used in the simulations is
$\Delta t=1~\mathrm{ns}$, which is much shorter than both the pulse
switching time and the molecular coherence time.
In addition, the excitation pulse generated by the RF signal generator
does not exhibit perfectly abrupt edges. The finite rise and fall times
of the electrical modulation, on the order of 20~ns, were therefore
included in the simulations. The incident field envelope is modeled as
\begin{equation}
E_{\mathrm{in}}(t)=E_0\,A(t),
\label{eq:supp_Ein}
\end{equation}
where $E_0$ is the constant field amplitude during the pulse and $A(t)$
is the temporal envelope of the excitation pulse, modeled using a
smooth square function based on hyperbolic tangent transitions
\begin{equation}
A(t)=
\frac{1}{2}\left(1+\tanh\left(\frac{t}{t_{\mathrm{rise}}}\right)\right)
\frac{1}{2}\left(1+\tanh\left(\frac{T_{\mathrm{on}}-t}{t_{\mathrm{fall}}}\right)\right),
\label{eq:supp_envelope}
\end{equation}
with $t_{\mathrm{rise}}\simeq t_{\mathrm{fall}}\simeq20$~ns and
$T_{\mathrm{on}}=1~\mu$s the pulse duration. This smooth switching
reproduces the finite response time of the electronic modulation chain.

Importantly, the detected field corresponds to the coherent sum of the
incident field and the field re-emitted by the molecular polarization
[Eq.~\eqref{eq:supp_thin}]. During the falling edge of the excitation
pulse, the incident field decreases rapidly while the molecular
polarization remains finite. The re-emitted field can therefore become
temporarily comparable in amplitude and opposite in phase to the
incident field, leading to a transient destructive interference. As a
result, the total detected field can pass close to zero before the
free-induction decay dominates. This feature, clearly visible in the
experimental traces, is directly reproduced by the numerical
simulations when realistic pulse switching times are included.

\subsection{Effective optical depth}
\label{sec:supp_odeff}

In practice, the experimental measurements were not always performed
exactly at the center frequency of the molecular transition. Small
frequency offsets between the excitation carrier and the line center
modify the effective interaction strength because both the absorptive
and dispersive parts of the molecular response vary rapidly near the
resonance.

For this reason, the effective optical depth used in the simulations
was treated as an adjustable parameter in order to reproduce the
experimental traces. The resulting values providing the best agreement
between simulations and measurements are summarized in
Table~\ref{tab:supp_odeff}.

\begin{table}[h]
\centering
\begin{tabular}{cc}
\hline
Pressure ($\mu$bar) & $\mathrm{OD}_{\mathrm{eff}}$ \\
\hline
19 & 0.52 \\
50 & 0.80 \\
100 & 1.10 \\
\hline
\end{tabular}
\caption{Effective optical depths used in the simulations.}
\label{tab:supp_odeff}
\end{table}

The ratio $\mathrm{OD}_{\mathrm{eff}}/\mathrm{OD}_0^{\mathrm{HITRAN}}$
is nearly constant across the three pressures ($0.25$, $0.23$, and
$0.26$ at 19, 50, and 100~$\mu$bar, respectively). This
pressure-independent scaling points to a common instrumental origin.
An error in the absolute calibration of the emitted field or of the
detection gain can be excluded: in the linear regime, the optical depth
is extracted from the relative attenuation and re-emission, from which
the incident field amplitude cancels out. The remaining candidates are
geometric, such as an effective interaction length shorter than the
geometric cell length or a partial overlap between the THz beam and the
absorbing gas column, or a non-absorbed component of the detected field
bypassing the gas---including possible direct electrical crosstalk from
the RF generator to the receiver chain, since
$f_{\mathrm{IF}}=f_{\mathrm{RF}}$.

\section{Signal-to-noise ratio and detection threshold}
\label{sec:supp_snr}

The detection sensitivity is characterized directly from the averaged
traces. The instrumental noise floor, measured on the empty-cell
baseline after the pulse, is $-38.1$~dBm, with an rms fluctuation of
$0.26$~dB, i.e. $5.9\,\%$ in relative power. This value is consistent
with the averaging of $M=1000$ sweeps of a noise correlated over
$N_c\approx2$ samples ($\sqrt{N_c/M}\approx5\,\%$), the correlation
length being estimated from the autocorrelation of the baseline. The
FID peak reaches $7.5$, $10.6$, and $11.8$~dB above the noise floor at
19, 50, and 100~$\mu$bar, respectively.

On the averaged trace, signal and noise powers are additive,
$\langle|E_s+n|^2\rangle=P_s+P_n$, and the floor fluctuations are
Gaussian to an excellent approximation (central-limit theorem), so
that the FID can be tested directly against the floor fluctuation
$\sigma_P$. At 19~$\mu$bar, the peak signal power exceeds the floor
fluctuation by a factor $P_s/\sigma_P\approx78$. In the linear
low-pressure regime, the FID field amplitude scales as the molecular
density, so that $P_s\propto p^2$ while the instrumental noise is
unchanged; a $3\sigma$ peak-detection criterion then extrapolates to a
minimum detectable pressure of
\begin{equation}
p_{\mathrm{min}}(3\sigma)=19\sqrt{3\,\sigma_P/P_s}
\approx3.7~\mu\mathrm{bar}.
\label{eq:supp_pmin}
\end{equation}

The threshold improves when the full temporal profile of the FID is
exploited. Correlating the recorded power trace with a smooth template
adjusted to the measured decay (matched filtering of the power
envelope, with the noise correlation length $N_c$ taken into account)
yields a detection statistic
\begin{equation}
D=\frac{1}{\sigma_P}\sqrt{\frac{1}{N_c}\sum_i P_s(t_i)^2}\approx660
\label{eq:supp_deflection}
\end{equation}
at 19~$\mu$bar, and, with $D\propto p^2$, a $3\sigma$ threshold of
\begin{equation}
p_{\mathrm{min}}(3\sigma)=19\sqrt{3/D}\approx1.3~\mu\mathrm{bar}
\end{equation}
under the present averaging conditions.